# Ultrasensitive and broadband MoS$_2$ photodetector driven by ferroelectrics

Xudong Wang[1,2][†], Peng Wang[1][†], Jianlu Wang[1]*, Weida Hu[1]*, Xiaohao Zhou[1], Nan Guo[1], Hai Huang[1], Shuo Sun[1], Hong Shen[1], Tie Lin[1], Minghua Tang[2], Lei Liao[3], Anquan Jiang[4], Jinglan Sun[1], Xiangjian Meng[1], Xiaoshuang Chen[1], Wei Lu[1], Junhao Chu[1]

[1]National Laboratory for Infrared Physics, Shanghai Institute of Technical Physics, Chinese Academy of Sciences, 500Yu Tian Road, Shanghai 200083, China

[2]School of Materials Science and Engineering, Xiangtan University, Xiangtan, Hunan 411105, China

[3]Department of Physics and Key Laboratory of Artificial Micro- and Nano-Structures of Ministry of Education, Wuhan University, Wuhan 430072, China

[4]Department of Microelectronics, Fudan University, 220 Handan Road, Shanghai 200433, China

Abstract: Photodetectors based on two dimensional (2D) materials have attracted growing interest. However, the sensitivity is still unsatisfactory even under high gate voltage. Here we demonstrate a $MoS_2$ photodetector with a poly(vinylidene fluoride-trifluoroethylene) ferroelectric layer in place of the oxide layer in a traditional field effect transistor. The dark current of the photodetector is strongly suppressed by ferroelectric polarization. A high detectivity (~$2.2\times10^{12}$ Jones) and photoresponsivity (2570 A/W) detector has been achieved under ZERO gate bias at a wavelength of 635 nm. Most strikingly, the band gap of few-layer $MoS_2$ can be tuned by the ultra-high electrostatic field from the ferroelectric polarization. With this characteristic, photoresponse wavelengths of the photodetector are extended into the near infrared (0.85-1.55 μm). A ferroelectrics/optoelectronics hybrid structure is an effective way to achieve high performance 2D electronic/optoelectronic devices.

During the past ten years, because of the unique properties of single layer graphene, two dimensional (2D) materials draw more and more attention for their potential applications in future nanoscale electronic/optoelectronic devices[1-13]. Molybdenum disulfide ($MoS_2$), consisting of layered S-Mo-S units structures bonded by van der Waals forces, have also been widely studied in recent years[2, 8, 9, 14-17]. $MoS_2$ is a typical semiconductor with a band gap range from 1.2 eV to 1.8 eV as the thickness decreases from bulk to monolayer[18-20]. Field effect transistors based on monolayer or multilayer $MoS_2$ possess high current ON/OFF ratios of up to $10^7$-$10^8$ [8, 21-23]. Owing to these special features, ultrathin $MoS_2$ becomes a promising candidate material for future electronic applications.

For transistors based on 2D graphene and $MoS_2$, traditional dielectric materials, such as $SiO_2$, $HfO_2$, and $Al_2O_3$, possessing linear dielectric response to an electric field, are usually selected as gate dielectrics[1, 8, 14, 16, 17]. Many interesting and meaningful physical properties have been discovered using these materials. Among these, the optoelectronic properties of a field effect transistor, especially the $MoS_2$ photodetector (sometimes named a phototransistor) have attracted intensive attention[14-17]. For these photodetectors, the photoresponsivity is relatively low, with the first monolayer $MoS_2$ phototransistors exhibiting a photoresponsivity of 7.5 mA/W[15]. The specific detectivity $D^*$ for a photodetector, a figure of merit used to characterize performance, is a metric of detector sensitivity. Among these photodetectors, the detectivity is unfavorable for practical applications. Additionally, for these traditional photodetectors, additional gate bias ($V_g$) and a large drain-source bias ($V_{sd}$) are essential for obtaining high sensitivity.

Such $V_g$ may induce a leakage between source and gate, and the large $V_{sd}$ leads to a significant increase in dark current between the source and drain, as well as a self-heating effect in the channel. These effects will not only cause a large power dissipation, but will also seriously degrade the performance, such as low sensitivity[24, 25]. In recent years, poly(vinylidene fluoride-trifluoroethylene) (P(VDF-TrFE)) ferroelectric polymer films and (lithium niobate) $LiNbO_3$ ferroelectric crystal has been used in nano electronic devices, such as nonvolatile memories[26-31]. In these devices, the function of the ferroelectric film is to tune the transport properties of the channel. However, ferroelectric materials combined with the photoelectric 2D materials have never been used for optoelectronic devices, for example photodetectors.

In this work, the $MoS_2$ transistor with a ferroelectric gate is used as a photodetector, wherein the few-layer $MoS_2$ serves as the photosensitive semiconducting channel while the remnant polarization of P(VDF-TrFE) is employed to depress the dark current of the $MoS_2$ semiconducting channel. The stable remnant polarization can provide an ultra-high local electrostatic field (~$10^9$ V/m within a several nanometer scale) in the semiconductor channel which is larger than that produced by gate bias in traditional field effect transistors[15-17]. With such an ultra-high electrostatic field, the few-layer $MoS_2$ channel is maintained in a fully depleted state, significantly increasing the sensitivity of the detector even at ZERO gate voltage. Based on these special properties, a photodetector with high detectivity ~$2.2\times10^{12}$ Jones and photoresponsivity up to 2570 A/W has been achieved. In addition, for the first time the photoresponse wavelengths of the ferroelectric polarization gating $MoS_2$ photodetector are extended from the visible to the near infrared (0.85-1.55 μm). Our MoS$_2$ photodetectors show comparable performances to those of commercial available silicon photodiodes (D$^*$~ 10$^{13}$ Jones and R ~300 A/W)[32].

A few-layer MoS$_2$ on a degenerately doped silicon substrate covered with a 285 nm thick silicon oxide was prepared using the Scotch® tape-based mechanical exfoliation method[2, 8]. The molecular configurations of MoS$_2$ are shown in Fig. 1a. The source and drain chromium (Cr, 5 nm)/gold (Au, 50 nm) electrodes were prepared using the lift-off method. The following step was coating the P(VDF-TrFE) (70:30 in mol%) film as the top gate. The molecular configurations of P(VDF-TrFE) are shown in Fig. 1b respectively. The ferroelectric dipole direction in the P(VDF-TrFE) chain is also shown in Fig. 1b. Then the ultrathin semi-transparent aluminum (~15 nm) electrodes were deposited. The thickness of the few-layer MoS$_2$ was confirmed by the Raman spectrum (Fig. S1). The peak location of $E_{2g}^1$ and $A_{1g}$ are 381.32 cm$^{-1}$ and 404.65 cm$^{-1}$, respectively, and the frequency difference between $E_{2g}^1$ and $A_{1g}$ vibration modes is 23.27 cm$^{-1}$. These data correspond to a film thickness of 2.1 nm for the triple-layer MoS$_2$[18, 19]. The optical image of the whole device based on MoS$_2$ and P(VDF-TrFE) is shown in Fig. 1c. The three-dimensional device structure schematic view of the ferroelectric polarization gating MoS$_2$ photodetector with laser beam illumination is shown in Fig. 1d.

Analysis continues with the ferroelectric properties of the P(VDF/TrFE) copolymer being characterized. A typical hysteresis loop for a P(VDF/TrFE) capacitor is shown in Fig. 2a. The thickness of the P(VDF/TrFE) layer is 300 nm. The coercive voltage is approximately 22.5 V and the remnant polarization value is 7 μC/cm$^2$. The polarization switching voltage can be reduced to approximately 5 V with decreasing thickness of the ferroelectric layer[27]. Next, the transfer curves $I_{sd}$-$V_{tg}$ (drain-source current $I_{sd}$ as a function of top gate voltage $V_{tg}$) of the MoS$_2$ transistor with ferroelectric polymer were investigated at room temperature (shown in Fig. 2b). The large memory window (~25 V) in *I-V* curves between the voltage rise and decrease is related to the ferroelectric polarization switching process. It is more obvious by comparing this curve with the one obtained with the SiO$_2$ back gate (inset of Fig. 2b). A mobility of $\mu \approx 86.5$ cm$^2$V$^{-1}$s$^{-1}$ is calculated from the polymer gated transfer curve using the method in Ref. [8]. The measured transfer characteristics exhibit a typical *n* type channel FET, showing good agreement with that of a conventional transistor[8]. With the back gate, a mobility of $\mu \approx 24.6$ cm$^2$V$^{-1}$s$^{-1}$ for the few-layer MoS$_2$ was obtained. The hysteresis behavior is related to the surrounding conditions or the charge transfer from neighboring adsorbates or charge injection into the trap sites on the substrate[33]. The difference in the mobility derived from top or back gates may be associated with the interface nature at the contacts[34, 35]. Based on the ferroelectric polymer/MoS$_2$ structure, we can achieve three different states: P(VDF/TrFE) without polarization (named as the "fresh" state), polarization up state (P$_{up}$) and polarization down state (P$_{down}$). The P$_{up}$ and P$_{down}$ states are achieved by poling P(VDF-TrFE) with -40 V and +40 V. The $V_{sd}$-$I_{sd}$ characteristics (without additional gate voltage and light illumination) of these three states are shown in Fig. 2c. In the $P_{up}$ state, $I_{sd}$ is the lowest compared to that of the other two states. This situation means that the depleted state of carriers in the MoS$_2$ channel is caused by the electrostatic field derived from the remnant polarization of P(VDF-TrFE). On the other hand, the $P_{down}$ state corresponds to the accumulated states of carriers in the MoS$_2$ channel. The cross sections of the device structures and equilibrium band diagrams at the different states are shown in Figs. 2d-2f. For a photodetector, the $V_{sd}$-$I_{sd}$ represents the dark current level without light illumination.

Next, the photoresponse of the MoS$_2$ photodetector was measured at a ZERO gate voltage. The photo switching properties (under $V_{sd} = 100$ mV) at a wavelength of 635 nm with the three states described above are shown in Fig. 3a. In the fresh state, the signal-to-noise-ratio, photoresponse current ($I_{ph}$) to dark current ($I_{dark}$) is very small. In the $P_{up}$ state, a signal-to-noise-ratio of $10^3$ is obtained, as can be seen from the photoresponse to pulsed laser illumination curves. In the $P_{down}$ state, $I_{ph}$ cannot be distinguished from the large $I_{dark}$ as the large thermionic and drift/diffusion currents dominate channel current. The results indicate that the electrostatic field produced by the surface charge at the domain surface of P(VDF-TrFE) is strong enough to cause total depletion/accumulation of carriers in the MoS$_2$ semiconducting channel. The domain direction in the ferroelectric material of a charge insufficiently compensated system can be altered by a depolarization field, which leads to the random distribution of ferroelectric domains and, eventually, to the malfunction of the device. (Indeed, before fabricating the ferroelectric layer, we have characterized the photoresponse properties of back gated MoS$_2$ photodetector, as shown in Figure S2. The performance of the traditional back-gate photodetector is very poor compared to the MoS$_2$ photodetector with ferroelectric polarization gating.) For the following measurement, the ferroelectric polarization in P(VDF-TrFE) was preset at a $P_{up}$ state by a short bias pulse of 40 V on the top gate. The $I_{sd}$-$V_{sd}$ for different illumination powers at a wavelength of 635 nm is shown in Fig. 3b. The $I_{sd}$-$V_{sd}$ curves, as shown in Fig. 3b, are linear and symmetric. Figure 3c shows the laser power dependence of $I_{ph}$ of the device with $V_{sd}$=5 V. The inset in Fig.3c is the ratio of $I_{ph}$ to $I_{dark}$ for light on and light off (dark states) at different illumination powers. It can be seen that the device is very sensitive to illumination power. The responsivity (*R*) of the photodetector is up to 2570 A/W and the detectivity is about ~2.2×10$^{12}$ Jones (under low illumination power 1 nW). The data are extracted from Fig. 3b, where $V_{sd}$ = 5 V and *P* = 1 nW, given by *R* = $I_{ph}$ / *P*, where $I_{ph}$ is the photo current in a detector and *P* is the illumination power[36, 37]. The value of *R* (2570 A/W) is larger than previously recorded result ~880 A/W (150 pW, $V_{sd}$=8 V, and $V_g$= -70 V)[8]. In addition, the detectivity, assuming that noise from dark current is the major factor, it is given by $D^* = RA^{1/2} / (2eI_d)^{1/2}$, where *R* is the responsivity, *A* is the area of the detector, *e* is the unit charge, and $I_d$ is the dark current[38]. The maximum detectivity ~2.2×10$^{12}$ Jones has been achieved, which is higher than that of detectors based on 2D materials. Note that, if a highly transparent electrode (for example graphene, Ag nano-wire) is chosen, a higher responsivity can be expected. Currently, the transmittance of the available semitransparent Al electrode is only 30% for our device. It suggests that a flawless device fabrication is needed for superior performance in the future.

In addition to the superior sensitivity and good photoresponsivity, we conducted time-resolved photoresponse experiments by periodically turning the illuminating laser on and off at a frequency of 0.5 Hz and recording the response signal with a high speed oscilloscope (shown in Fig. S3). Fig. 3e shows a complete on/off cycle in which the photocurrent exhibits rapid rise/fall and reaches a steady saturation. The rising and falling edges of the response current are perfectly fitted by a single exponential function. The rise ($\tau_r$) and decay ($\tau_f$) times of the photocurrent, are ~1.8 ms and ~2 ms, respectively, which is relatively fast for state-of-the-art 2D photoconductive photodetectors. In our case, the improvement of response time may result from the interface of P(VDF-TrFE) and MoS$_2$, the surface trap state of MoS$_2$ may be encapsulated or passivated by the fluorine or hydrogen atoms from polarized P(VDF-TrFE). Furthermore, the signals recorded by the oscilloscope remain nearly unchanged after 90, 000 cycles of operation (as shown in Fig. 3f), pointing to the excellent stability and reliability of the photodetectors. The stability of the photo-switching behavior is related to the ferroelectric polarization stability of the P(VDF-TrFE) top gate[39, 40] and the reliability of the P(VDF-TrFE)-passivated MoS$_2$ channel.

All the prominent properties of our photodetector benefit from the ferroelectric-polarization-induced ultra-high electrostatic field of the P(VDF-TrFE). The local electric field at the interface between P(VDF-TrFE) and MoS$_2$ layers can be estimated from $\sigma=\varepsilon\varepsilon_0 E$, where $\sigma$ is the charge density at the surface of P(VDF-TrFE) film, which is related to the remnant polarization ($P_r$) of the ferroelectric materials[27], $\varepsilon$ is the dielectric constant of material (the dielectric constant of MoS$_2$ is approximately 4-6[41, 42], $\varepsilon_0$ is the vacuum permittivity, $E$ is the electric field strength, and $P_r$ is approximately 7.0 μC/cm$^2$ calibrated from Fig. 2a. The calculated electric field applied to the triple-layer MoS$_2$ is about $0.5 \times 10^9$ V/m. It is very difficult to obtain such an ultra-high electric field for conventional FETs as described by several research groups[27, 43, 44].

The typical photoresponse spectra of a MoS$_2$ photodetector is from visible to near infrared (0.85 μm) as the band gap of MoS$_2$ ranges from 1.2 eV to 1.8 eV. As reported by Oriol Lopez-Sanchez *et al.*, photoresponse is negligible at wavelengths longer than 680 nm (corresponding to one photon energy 1.8 eV) for monolayer MoS$_2$[14]. For triple-layer MoS$_2$, the broad absorption tail corresponding to the indirect band transition extends to near 0.85 μm, which is in good agreement with the theoretical prediction[17]. In this work, we also characterized the photoresponse properties of fresh state MoS$_2$ photodetector, where the response wavelength ends at ~900 nm (shown in Fig. S4). A wider spectral response for a photodetector is important and meaningful. Therefore the response of the detector to light with long wavelength was investigated. The photoresponse of the detector to laser illumination with different wavelengths is shown in Fig. 4a. Surprisingly, it is found that there is still appreciable optical current response to light with wavelength up to 1.55 μm. For the first time, our work broadens the detection of a MoS$_2$ photodetector from 0.85 μm to 1.55 μm.

To confirm the effect of the electric field polarization on the band structure of few-layer MoS$_2$, we carried out micro-photoluminescence (PL) measurements on the fresh (ferroelectric domain random) and poled (domain aligned) samples (in Fig. 4b). The PL emission for the pole sample is red-shifted compared to that of the fresh sample, In addition, the PL emission intensity is also increased after the cover of poled P(VDF-TrFE). This effect may be related to the electric field modifying the band structure of the MoS$_2$ and in turn increasing the density of states of the carriers. With the increase in the carrier population, the radiative recombination of carriers increases, which leads to an increase in PL intensity[45, 46]. Similarly, the band structure of MoS$_2$ can be tuned by the external strain and confirmed by PL spectrum[45-48]. Note that an optical bandgap measured by PL is different from the bandgap of electron transport due to the exciton binding energy. It has been theoretically predicted that the energy gap of the bi-layer MoS$_2$ can be tuned by the external electric field[49, 50]. The change in the band-gap in triple-layer MoS$_2$ under external electric fields applied perpendicular to the layers was also calculated using the density functional theory. The band structure of triple-layer MoS$_2$ as a function of applied different external electric field is shown in Figs. 4c-e. For example, a 0.40 V/nm of electric field strength can reduce the band gap from 1.09 to 0.71 eV. (Details of the calculation are shown in supplementary information). The indirect bandgap is reduced with an increase of the applied electric field, and the relationship between $E_g$ and electric field is linear (as shown in Fig. S6).

Further studies are needed to clarify the fundamental physics principles of the ferroelectric polarization tuning for the bandgap of few-layer MoS$_2$. Nevertheless, the ferroelectric/MoS$_2$ hybrid structure photodetector shows an outstanding performance for detection. Similar electric and photoresponse properties have been also achieved on the same structure with 4-layer MoS$_2$ photodetector (shown in Fig. S7 and S8). Our photodetector shows a high sensitivity under weak illumination, such as fluorescent lamp (Mov. S1), demonstrating that the few-layer MoS$_2$ is very promising for a visible to near-infrared digital camera.

In summary, we have fabricated the ferroelectric polymer film gated triple-layer MoS$_2$. The MoS$_2$ device exhibited outstanding photodetection capabilities compared to traditional MoS$_2$ FET photodetectors. The device exhibits a maximum attainable photoresponsivity of 2570 A/W, and high detectivity of $2.2\times10^{12}$ Jones, which are record values compared to the previous reports on MoS$_2$ and other 2D material photodetector. In addition, the broad spectral regions detection, stable and fast photoresponse of the detector are also superior to recently reported MoS$_2$ photodetectors. The infrared (0.85-1.55 μm) photoresponse of the ferroelectric polarization gating MoS$_2$ photodetector suggests that such a MoS$_2$ photodetector is very promising for use in optical communications. The energy gap engineering of a 2D material system combined with an ultra-high ferroelectric-polarization-induced electrostatic field is an attractive research area for next-generation high performance 2D electronic/optoelectronic devices. There are many fruitful and fundamental physics issues in ferroelectric/photoelectric 2D material hybrid systems to be explored in future studies.

**Fabrication of the MoS$_2$ photodetectors.**

All measurements were performed under ambient conditions. Triple layers of MoS$_2$ were exfoliated from commercially available crystals of molybdenite (SPI Supplies brand Moly Disulfide) using the Scotch®-tape micromechanical cleavage technique pioneered for the production of graphene. Triple layers of MoS$_2$ were deposited on a ~285 nm thick SiO$_2$ dielectric layer on top of a highly-doped p-type Si wafer (resistivity<5×10$^{-3}$ Ω·cm). Electrical contacts were patterned on top of MoS$_2$ flakes using a conventional lift-off technique. Cr (5 nm) and Au (50 nm) electrodes were deposited by thermal evaporation at room temperature. The device was then annealed at 200 °C in a vacuum tube furnace for 2 hours (100 sccm Ar) to remove resist residue and to decrease contact resistance. Then the top gate P(VDF-TrFE) (70:30 mol%) films were prepared by spin coating on the top of the MoS$_2$. Finally, ultrathin aluminum films were deposited by thermal evaporation as the top gate semi-transparent electrodes.

**Characterization of the photodetectors**

We performed the electrical and optoelectrical characterization of our device at room temperature using a Lake Shore probe station and Agilent semiconductor parameter analyzer.

In this work, electrical measurements were carried out using an Agilent B2902A, Signal Recovery current preamplifier, lock-in amplifier Signal Recovery 7270 and a current preamplifier Signal Recovery 5182 on the Lake Shore probe station. The photoresponse to laser excitation used a focused $\lambda$=500-1550 nm laser beam. A monochrometer was used for wavelength-dependent measurements of the photocurrent. The Raman spectra and PL spectra were acquired by the Lab Ram HR800 from HORIBA (excitation wavelength 532 nm, power 2 mW, spot size 2-3 μm). Time-resolved photoresponse were achieved by high speed the Tektronix MDO3014 Oscilloscope.

**Acknowledgements**

This work was partially supported by the Major State Basic Research Development Program (Grant No. 2013CB922302 and 2014CB921600), Natural Science Foundation of China (Grant Nos. 11374320, 11322441 and 61440063), and Fund of Shanghai Science and Technology Foundation (Grant Nos. 14JC1406400).

**Author contributions**

J.W and W.H conceived and supervised the research. X.W and J.W. fabricated the devices. X.W and P.W. performed the measurements. X.Z carried out the calculated part. J.W and W.H wrote the paper. All authors discussed the results and revised the manuscript.

**Additional information**

Supplementary information is available in the online version of the paper. Correspondence and requests for materials should be addressed to J.W. and W.H.

**Competing financial interests**

The authors declare no competing financial interests.

**Figures**

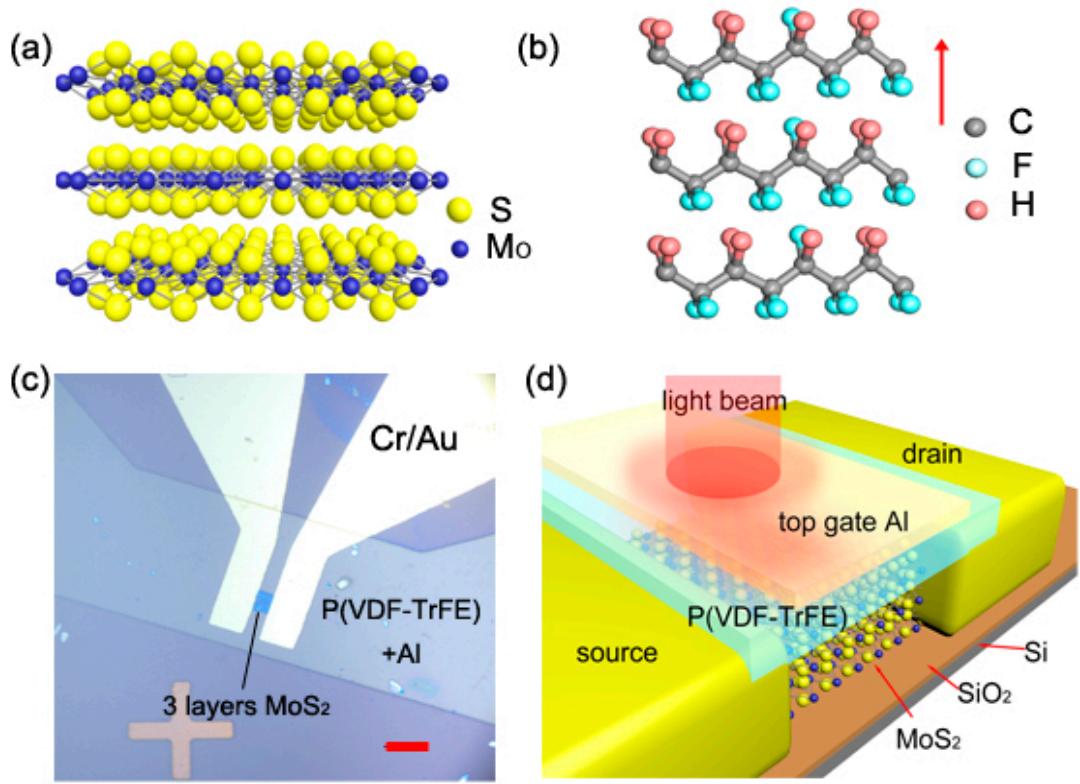

**Figure 1. Fabrication and structure of few-layer MoS$_2$ photodetector. a**, Schematic structure of triple-layer MoS$_2$. **b**, Schematic structure of P(VDF-TrFE) ferroelectric polymer. The polarization direction of the polymer is pointed out by arrow. **c**, Optical image of the whole device. The device is comprised of triple-layer MoS$_2$ with Cr/Au contract, 250 nm P(VDF-TrFE) ferroelectric polymer and semi-transparent aluminum top electrode. Scale bar, 10μm.   **d**, Three dimensional schematic view of the triple-layer MoS$_2$ photodetector with monochromatic light beam.

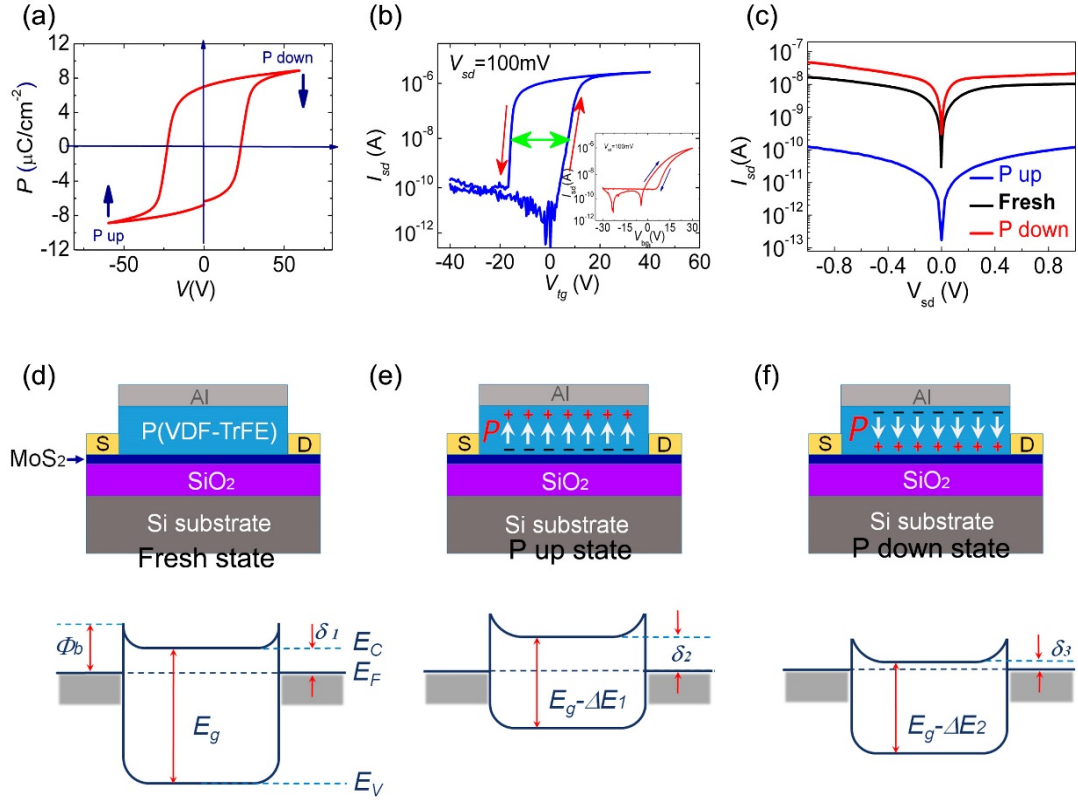

**Figure 2. Ferroelectric polarization related electric properties of the P(VDF-TrFE)/MoS$_2$ hybrid structure. a**, The ferroelectric hysteresis loop 300 nm P(VDF-TrFE) film capacitor. It is measured using Sawyer-Tower circuit at 1 Hz applied voltage frequency. **b**, The transfer curves of triple-layer MoS$_2$ channel with P(VDF-TrFE) ferroelectric polymer gate on dark state at room temperature. The transfer characteristics of triple-layer MoS$_2$ with SiO$_2$ back gate are shown in the inset. **c**, The $V_{sd}$-$I_{sd}$ characteristics (at ZERO gate voltage) with three states of ferroelectric layer. The three states are: fresh state (ferroelectric layer without polarization), polarization up ("P up", polarized by a pulse $V_g$ of -40 V) and polarization down ("P down", polarized by a pulse $V_g$ of -40 V) states, respectively. **d-f**, The cross section structures of the device and equilibrium energy band diagrams of three different ferroelectric polarization states with $V_{sd}$=0 V. For the band diagrams of different states, small

Schottky barriers at the source and drain electrodes are considered. $E_F$, $E_c$ $E_V$, $\Phi_B$, and $E_g$ are the Fermi level energy, minimum conduction band energy, maximum valence band energy, Schottky barrier height and bandgap of $MoS_2$, respectively. $\delta$ is the height from bottom of conduction band to the Fermi level. $\delta_1$, $\delta_2$ and $\delta_3$ are related to the three states.

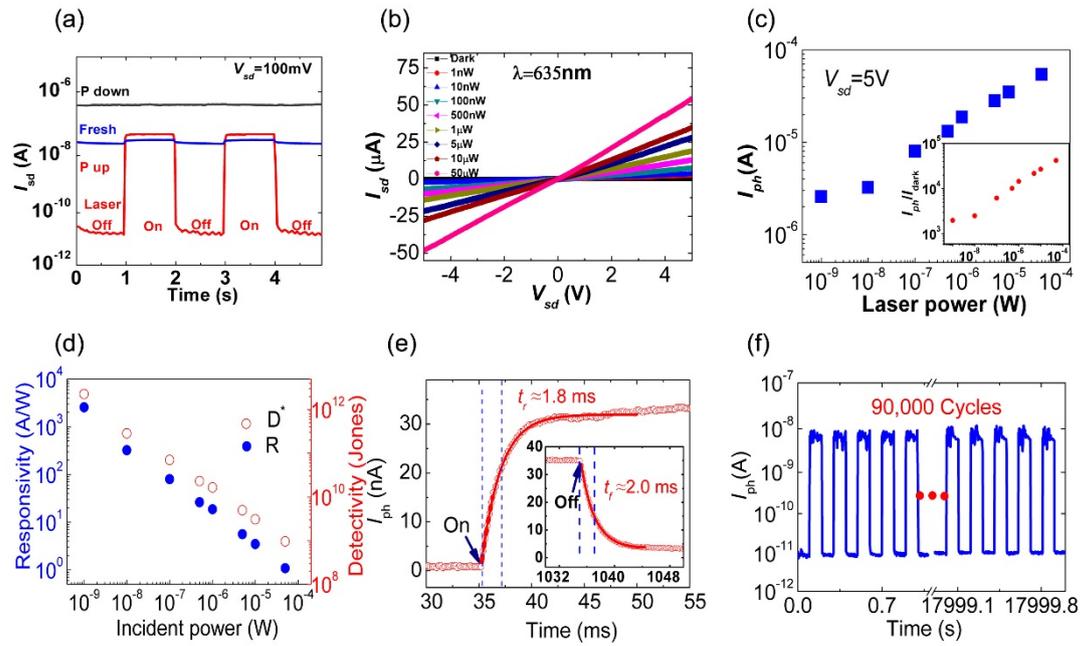

**Figure 3. Photoresponse properties of the ferroelectric polarization gating triple-layer MoS$_2$ photodetector. a,** Photoswitching behavior of ferroelectric polarization gating triple-layer MoS$_2$ photodetector at three states ($\lambda$=635 nm, $V_{sd}$=100 mV, $P$=100 nW). **b,** Drain-source characteristic of the photodetector in the dark and under different illuminating light powers (1nW to 50 μW). **c,** Dependence of photocurrent on illumination powers. Inset is the ratio of photoresponse current to dark current under different illumination powers. **d,** Photoresponsivity of the MoS$_2$ phototransistor, showing high sensitivity. The device exhibits a photoresponsivity of 2570 A/W for an laser power of 1 nW, and the detectivity is up to ~2.2×10$^{12}$ Jones. **e,** The rise and fall of the photocurrent and the fitted data using exponential functions (recorded by $V_{sd}$=100 mV and $P$=100 nW). **f,** Photocurrent response during 90,000 cycles of operation at $V_{sd}$=100 mV and $P$=100 nW. The device shows an endurable photoresponse.

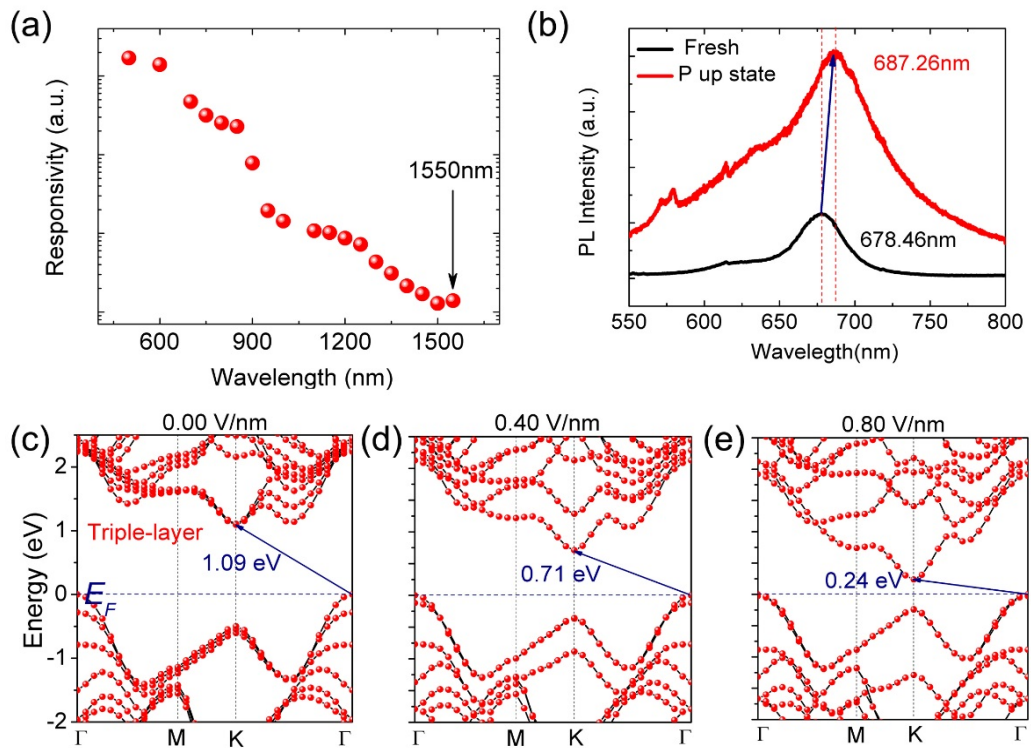

**Figure 4. Infrared photoresponse of ferroelectric polarization gating few-layer MoS$_2$ photodetector beyond its intrinsic bandgap. a,** Photoresponsivity of a similar polarization-gating triple layer MoS$_2$ photodetector as a function of light wavelength from 500 nm to 1550 nm at $V_{sd}$=1 V and $P$=100 μW). **b**, Comparison photoluminescence spectrum of fresh and polarization gating triple-layer MoS$_2$. **c-e**, The band structures evolution of triple-layer MoS$_2$ under different external electric field (0.0, 0.4, and 0.8 V/nm) by first-principle DFT calculations.

# Supplementary Information for

# Ultrasensitive and broadband MoS$_2$ photodetector driven by ferroelectrics

Xudong Wang[1,2][†], Peng Wang[1][†], Jianlu Wang[1]*, Weida Hu[1]*, Xiaohao Zhou[1], Nan Guo[1], Hai Huang[1], Shuo Sun[1], Hong Shen[1], Tie Lin[1], Minghua Tang[2], Lei Liao[3], Anquan Jiang[4], Jinglan Sun[1], Xiangjian Meng[1], Xiaoshuang Chen[1], Wei Lu[1], Junhao Chu[1]

[1]National Laboratory for Infrared Physics, Shanghai Institute of Technical Physics, Chinese Academy of Sciences, 500Yu Tian Road, Shanghai 200083, China

[2]School of Materials Science and Engineering, Xiangtan University, Xiangtan, Hunan 411105, China

[3]Department of Physics and Key Laboratory of Artificial Micro- and Nano-Structures of Ministry of Education, Wuhan University, Wuhan 430072, China

[4]Department of Microelectronics, Fudan University, 220 Handan Road, Shanghai 200433, China

Email: jlwang@mail.sitp.ac.cn, wdhu@mail.sitp.ac.cn

**Contents**

## 1. The Raman spectra of triple layer MoS$_2$.

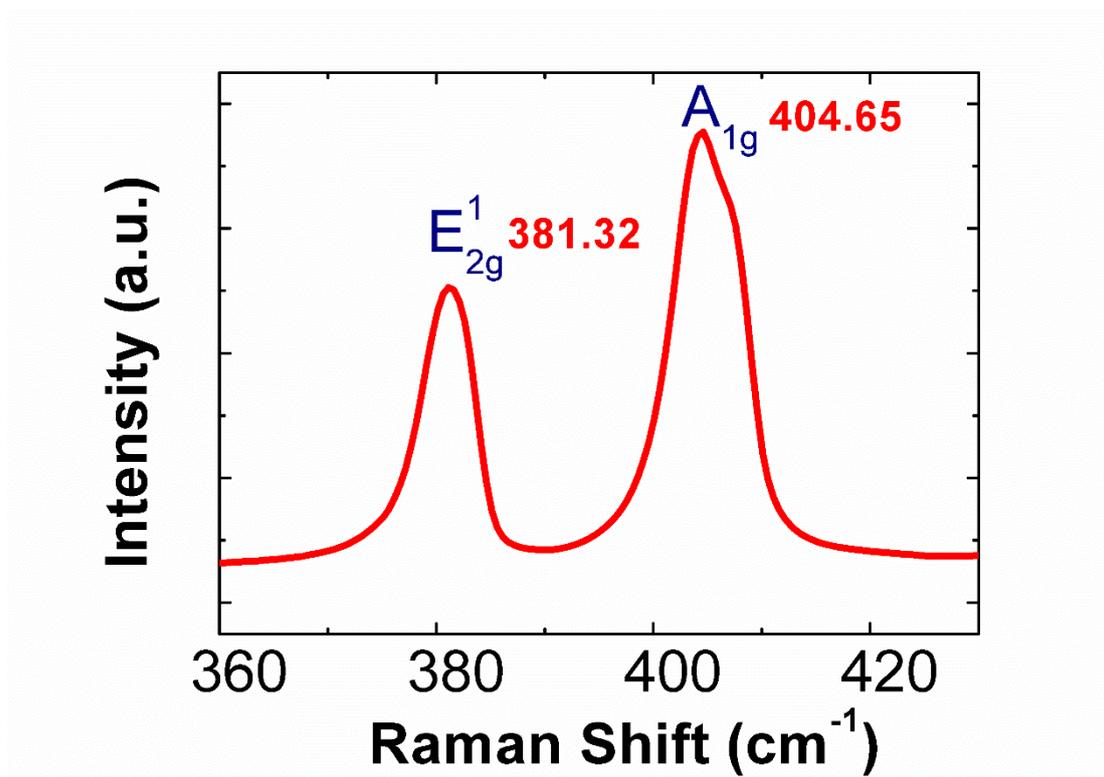

Figure S1. The Raman spectra of triple layer MoS$_2$. The peak position of $E^1_{2g}$ and $A_{1g}$ are 381.32 cm$^{-1}$ and 404.65 cm$^{-1}$, respectively. The separation between $E^1_{2g}$ and $A_{1g}$ modes is approximately 23.27 cm$^{-1}$.

## 2. Photoresponse properties of back-gated triple-layer MoS$_2$ photodetector.

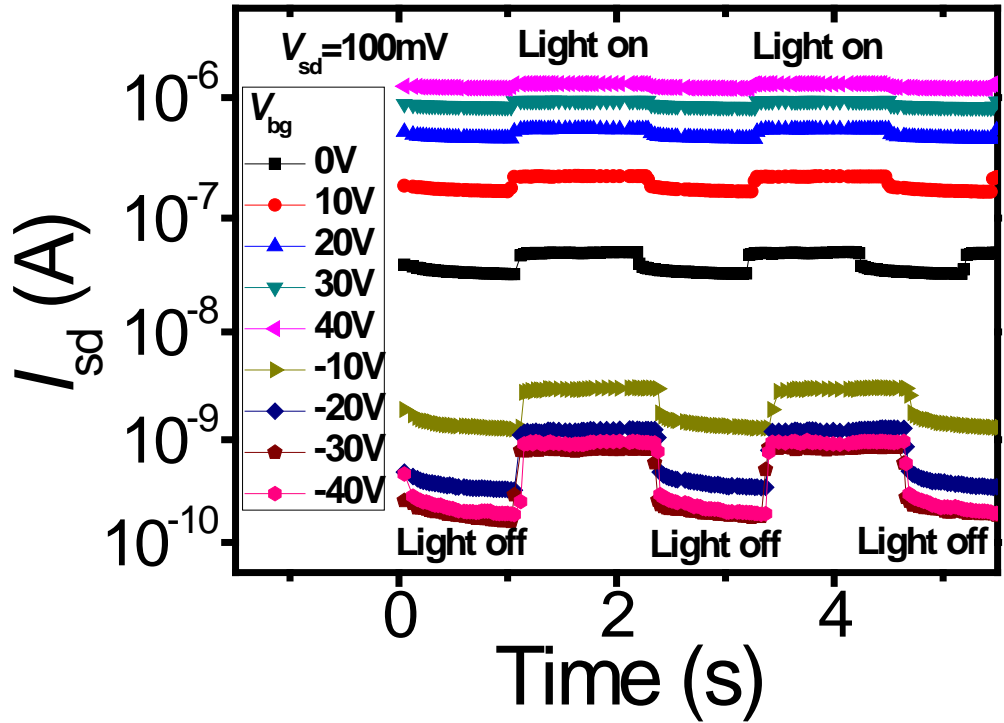

Figure S2. Photoresponse behavior of back-gated triple-layer MoS$_2$ photodetector with different back gate voltages and $V_{sd}$=100mV at a wavelength of 635 nm ($P$=100nW).

## 3. Time-resolved photoresponse of ferroelectric polarization gating MoS$_2$ photodetector

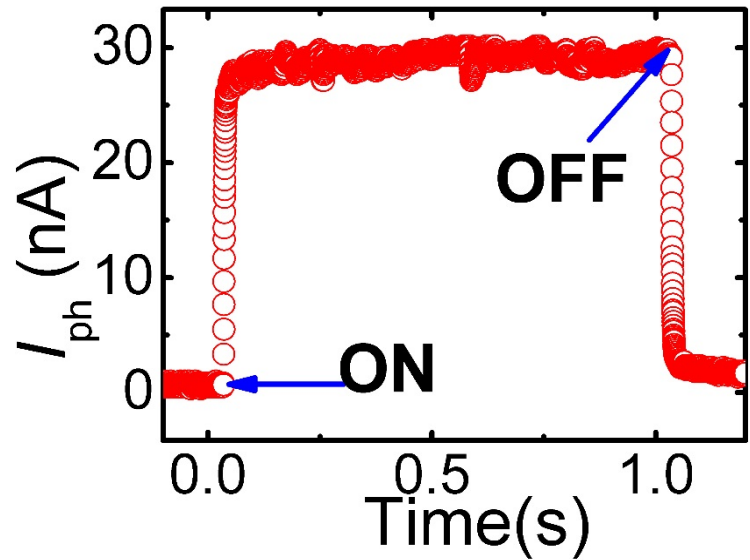

Figure S3. Time-resolved photoresponse of ferroelectric polarization gating MoS$_2$ photodetector. The data are recorded by a high speed oscilloscope at $V_{sd}$=100mV and $P$=100nW.

## 4. Photoresponse of tri-layer MoS₂ photodetector on fresh state

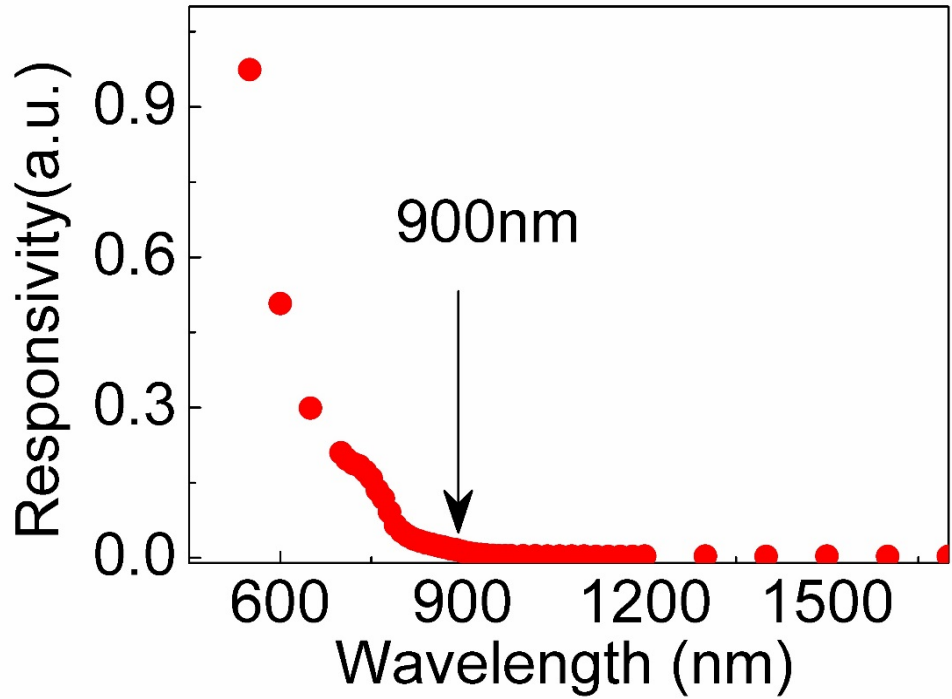

Figure S4. Photoresponsivity of triple-layer MoS₂ device (without ferroelectric) as a function of illumination wavelength. The device shows a decreasing responsivity as the illumination wavelength changes from 500 nm to 900 nm. When the wavelength is larger than 900, there is nearly no photoresponse for the device.

## 5. Calculations on band structures evolution of triple-layer MoS$_2$

The first-principle DFT calculations were performed within the DMol$^3$ code[S1]. The generalized gradient approximation (GGA) with the Perdew–Burke–Ernzerhof (PBE) function was utilized as the exchange–correlation function[S2]. The DFT-D (D stands for dispersion) approach within the Grimme scheme was adopted for the van der Waals corrections[S3]. DFT Semi-core Pseudopots (DSPP), which induce some degree of relativistic correction into the core, were used for the core treatment. Moreover, double numerical atomic orbital plus polarization was chosen as the basis set, with the global orbital cutoff of 4.6 Å The k-point was set to 7×7×1 for the structural optimization and 11×11×1 for the electronic properties calculations, and the smearing value was 0.005 Ha. The convergence tolerances of energy, maximum force, and maximum displacement were set to $1.0×10^{-5}$ Ha, 0.002 Ha/Å and 0.005 Å respectively. The vacuum length between two adjacent images in the supercell is longer than 15 Å. The external electric field (0 to 1.0 V/nm) was applied in the direction perpendicular to the triple-layer plane, and then geometry relaxation was carried out.

Furthermore, the band gap ($E_g$) value as a function of electric field is displayed in Figure S6. Clearly. $E_g$ is linearly dependent on electric field with a slope of 1.12. The linear relationship of $E_g$ and electric field was also reported in the transition-metal dichalcogenide bilayers under external electric field[S4].

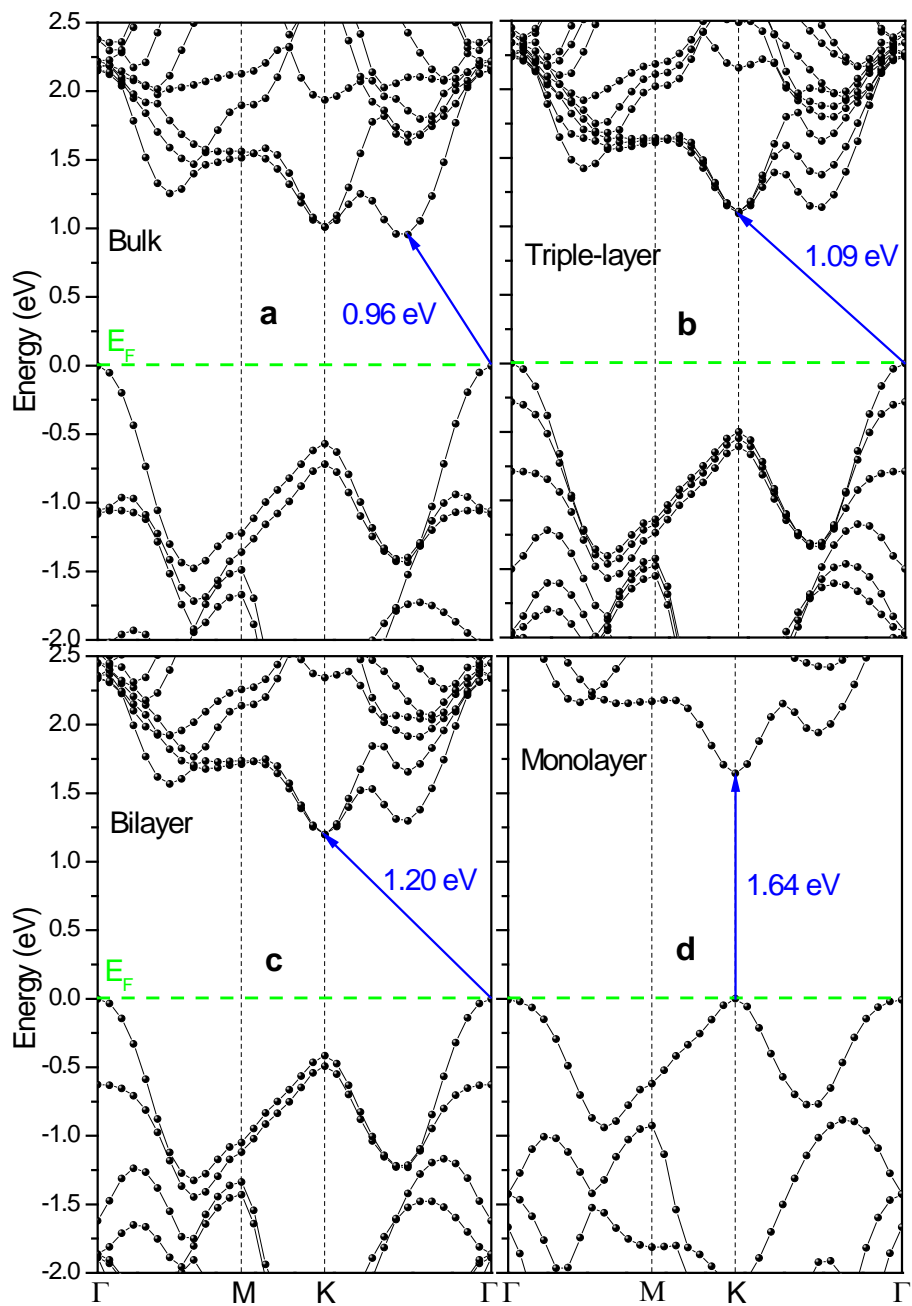

Figure S5. Computed electronic band structures of bulk (a), triple-layer (b), bilayer, and monolayer (d) MoS$_2$. The Fermi level ($E_F$) is shifted to the top of the valence band.

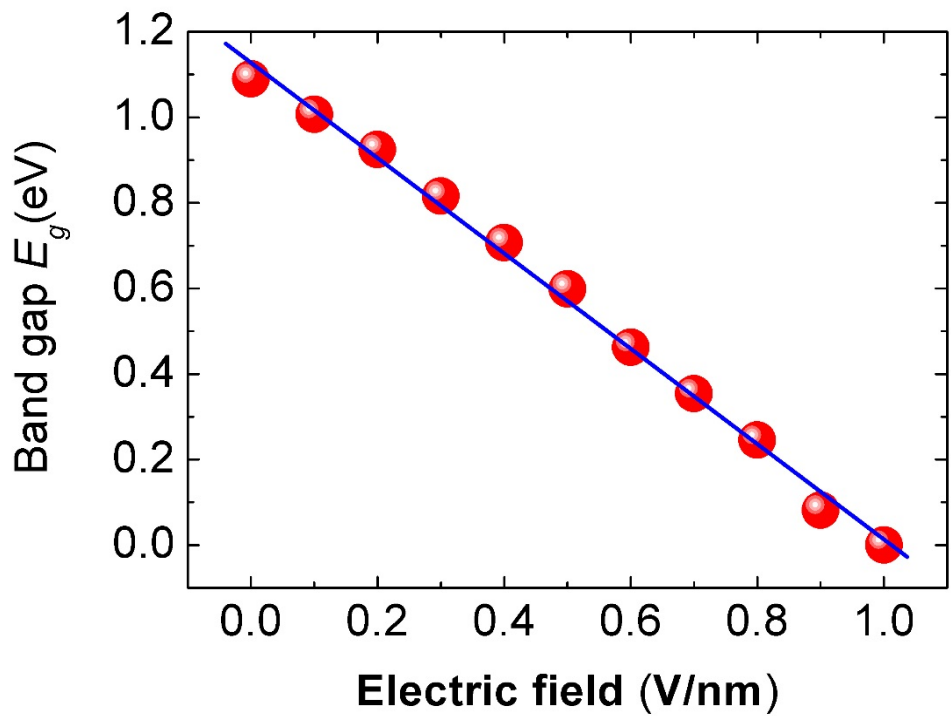

Figure S6. Band gap versus applied electric field for triple-layer MoS$_2$. The red (dotted) and blue (solid line) are the calculated data and fitted line, respectively.

# 6. Fabrication processing, structure and photoresponse properties of four-layer MoS₂ ferroelectric field effect transistor.

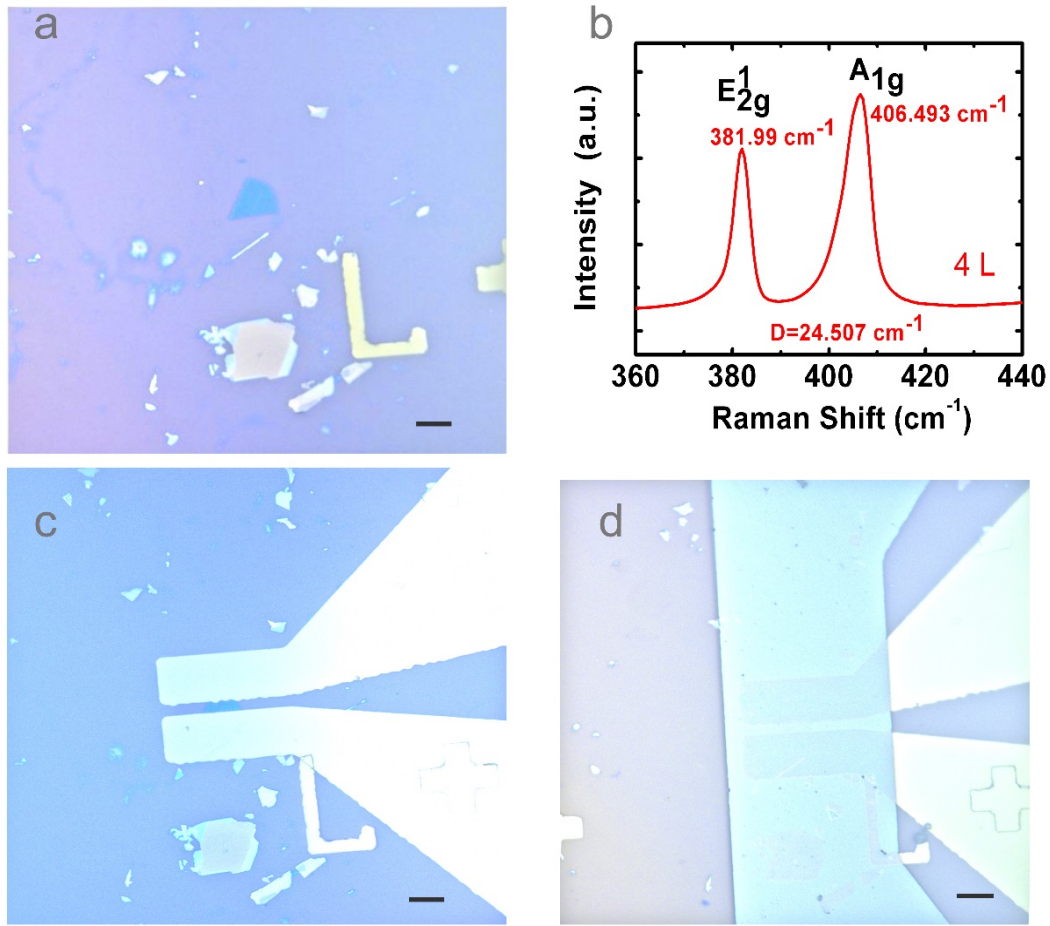

Figure S7. Fabrication and structure of four-layer MoS$_2$ detector. **a**, Optical image of the four-layer MoS$_2$. Scale bar, 10μm. **b**, The Raman spectra of the four-layer MoS$_2$. The peak position of the $E_{2g}^1$ and $A_{1g}$ are 381.99 cm$^{-1}$ and 406.49 cm$^{-1}$, respectively. The separation between $E_{2g}^1$ and $A_{1g}$ modes is approximately 24.50 cm$^{-1}$. **c**, Optical image of the four-layer MoS$_2$ on top of a SiO$_2$-silicon substrate with source and drain (chromium/gold, Cr/Au) electrodes. Scale bar, 10μm. **d**, Optical image of the complete device based on **c**. The device is comprised of four-layer MoS$_2$ with Cr/Au

contract, 300 nm P(VDF-TrFE) ferroelectric polymer and semi-transparent aluminum top electrode. Scale bar, 10μm.

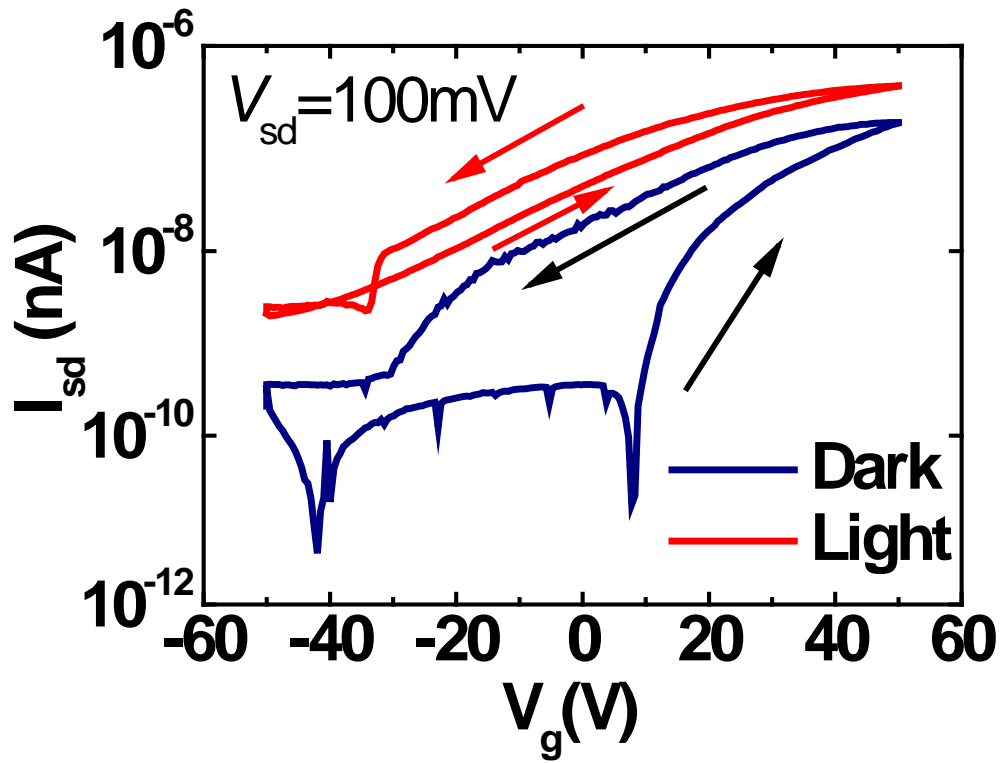

Figure S8. The transfer curves of four-layer $MoS_2$ photodetector in dark (red line) and light (blue line) states, respectively. The wavelength of the illumination light is 635nm.

# 7. Photoresponse properties of ferroelectric polarization gating MoS$_2$ photodetector in the infrared wavelengths

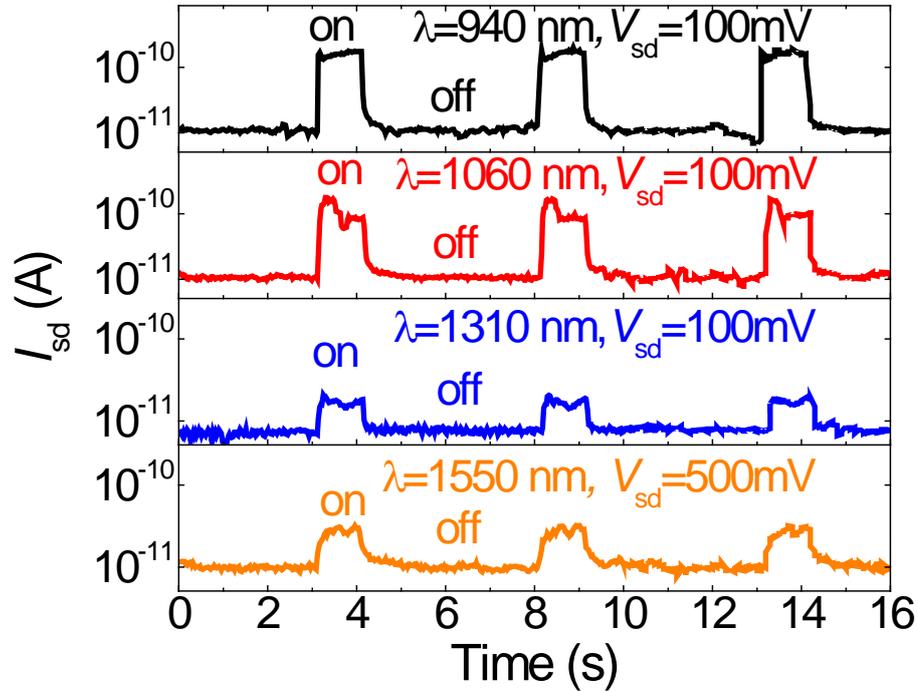

Figure S9. Infrared photoswitching behavior of ferroelectric polarization gating few-layer MoS$_2$ photodetector under 960 nm, 1060 nm, 1310 nm and 1550 nm wavelength laser illuminations, respectively.

# Movies

Mov S1

Photoresponse behavior of ferroelectric polarization gated few-layer $MoS_2$ photodetector under fluorescent lamp illumination.